%% file: root.tex
\documentclass[letterpaper, 10 pt, conference]{ieeeconf}  % Comment this line out if you need a4paper

\IEEEoverridecommandlockouts                              % This command is only needed if 
\usepackage{blindtext}
\usepackage{graphicx}
\usepackage[utf8]{inputenc}
\usepackage{graphics} % for pdf, bitmapped graphics files
\usepackage{epsfig} % for postscript graphics files
\usepackage{mathptmx} % assumes new font selection scheme installed
\usepackage{amsmath} % assumes amsmath package installed
\usepackage{amssymb}  % assumes amsmath package installed
\usepackage{cite}
\usepackage{physics}
\usepackage{multicol}
\usepackage{mathtools, cuted}
\usepackage[cmintegrals]{newtxmath}
\usepackage{epstopdf}
\usepackage{graphics} % for pdf, bitmapped graphics files
\usepackage{epsfig} 

\usepackage{mathptmx} % assumes new font selection scheme installed
\usepackage{amsmath} % assumes amsmath package installed
\usepackage{amssymb}  % assumes amsmath package installed
\usepackage{cite}
\usepackage{multicol}
\usepackage{mathtools, cuted}
\usepackage[cmintegrals]{newtxmath}
\usepackage{mathtools}

\usepackage{mathptmx}
\usepackage{helvet}
\usepackage{courier}
\usepackage{subfigure}
\usepackage{type1cm}
\usepackage{titlesec}
\usepackage{epsfig} % for postscript graphics files
\usepackage{mathptmx} % assumes new font selection scheme installed
\usepackage{amsmath} % assumes amsmath package installed
\usepackage{amssymb}  % assumes amsmath package installed
\usepackage{cite}
\usepackage{multicol}
\usepackage{mathtools, cuted}
\usepackage[cmintegrals]{newtxmath}
\usepackage{makeidx}         % allows index generation
\usepackage{algpseudocode}% http://ctan.org/pkg/algorithmicx

\usepackage{multicol}        % used for the two-column index
\usepackage[bottom]{footmisc}% places footnotes at page bottom
\usepackage{caption}
\usepackage{nomencl}
\usepackage[usenames, dvipsnames]{color}
\usepackage{url}
\usepackage{siunitx}
\usepackage{algorithm2e}

\usepackage{multicol}        % used for the two-column index
\usepackage[bottom]{footmisc}% places footnotes at page bottom
\usepackage{caption}
\usepackage{titlesec}
\usepackage{nomencl}
\usepackage[usenames, dvipsnames]{color}
\usepackage{multirow}
\usepackage{adjustbox}
\usepackage{amsthm}
 
\theoremstyle{definition}

\theoremstyle{remark}

\newcommand{\myfloor}[1]{\left \lfloor #1 \right \rfloor} 

\usepackage{threeparttable}
\usepackage{booktabs}
\usepackage{stmaryrd}

\title{\LARGE \bf
Can a Laplace PDE Define Air Corridors through Low-Altitude Airspace?}

\author{Aeris El Asslouj$^{1}$, Ella Atkins$^{2}$, and  Hossein Rastgoftar$^{1,3}$% <-this % stops a space
\thanks{*This work has been supported by the National Science Foundation under Award Nos. 2133690 and 1914581.}% <-this % stops a space
\thanks{$^{1}$Electrical and Computer Engineering Department, University of Arizona, Tucson, Arizona, USA {\tt\small aymaneelasslouj@arizona.edu}}
\thanks{$^{2}$Aerospace and Ocean Engineering Department, Viginia Tech, Blacksburg, Virginia, USA
24061 {\tt\small ematkins@vt.edu}}%
\thanks{$^{3}$Aerospace and Mechanical Engineering Department, University of Arizona, Tucson, Arizona, USA {\tt\small hrastgoftar@arizona.edu}}%
}

\begin{document}

\maketitle
\thispagestyle{empty}
\pagestyle{empty}

%%%%%%%%%%%%%%%%%%%%%%%%%%%%%%%%%%%%%%%%%%%%%%%%%%%%%%%%%%%%%%%%%%%%%%%%%%%%%%%%
\begin{abstract}
Urban Uncrewed Aircraft System (UAS) flight will require new regulations that assure safety and accommodate unprecedented traffic density levels.  Multi-UAS coordination is essential to both objectives.  This paper models UAS coordination as an ideal fluid flow with a stream field  governed by the Laplace partial differential equation. Streamlines spatially define closely-spaced deconflicted routes through the airspace and define air corridors that safely wrap  buildings and other structures so UAS can avoid collision even when flying among low-altitude vertical obstacles and near mountainous terrain.  We divide a city into zones, with each zone having its own sub-network, to allow for modularity and assure computation time for route generation is linear as a function of total area. We  demonstrate the strength of our proposed approach by computing air corridors through low altitude airspace of select cities with tall buildings. For US cities, we use open LiDAR elevation data to determine surface elevation maps. We select non-US cities with existing high-fidelity three-dimensional landscape models. 
% Our air networks are interconnected layers of air corridors generated based on ideal fluid flow through solving the Laplace PDE. Air corridors  We subdivide each city into zones with each zone having its own sub-network to allow for modularity and ensure computation time for generation is linear as a function of total area. We also present sample air networks for major cities around the world which demonstrate that our method is successful at generating dense connected air networks. 
% To accommodate increased urban drone usage, cities have to provide an organizational infrastructure to control UAS traffic for maximum airspace usage. We propose a fully automatic method to generate urban air networks for low-altitude flights analogous to ground road networks. For US cities, we use LiDAR elevation data from the USGS to determine their surface elevation maps. For non-US cities, we use existing high-fidelity 3d models. Our air networks are interconnected layers of air corridors generated based on ideal fluid flow through solving the Laplace PDE. Air corridors wrap around buildings and other structures so drones can avoid static obstacles simply by staying within the network. We subdivide each city into zones with each zone having its own sub-network to allow for modularity and ensure computation time for generation is linear as a function of total area. We also present sample air networks for major cities around the world which demonstrate that our method is successful at generating dense connected air networks. 
\end{abstract}

%%%%%%%%%%%%%%%%%%%%%%%%%%%%%%%%%%%%%%%%%%%%%%%%%%%%%%%%%%%%%%%%%%%%%%%%%%%%%%%%
\input{introduction}

\input{methodology}

\input{results}

\input{conclusion}

\bibliographystyle{IEEEtran}
\bibliography{sample.bib}

\end{document}

%% file: introduction.tex
\section{Introduction}

The number of Uncrewed Aircraft Systems (UAS) flying over and within urban regions is projected to rapidly increase \cite{mordorintelligence} for applications such as package delivery \cite{jung2017analysis}, aerial photography \cite{eckert2019using}, and airborne sensing \cite{savkin2019method}. For cities to safely accommodate all new UAS operations, higher air traffic densities, dynamic routing, and UAS-to-UAS coordination must all be supported. UAS will likely operate primarily at low altitudes despite a complex urban landscape to assure separation from new Advanced Air Mobility (AAM) traffic and to maximize efficiency in short-hop ($<2$ km) flights.

Increased aerial traffic densities suggest an increased risk of collisions. Apart from Automatic Dependent Surveillance Broadcast (ADS-B), which does not scale, there is currently no capability for UAS-to-UAS detect-and-avoid (DAA) or separation assurance through cooperative routing. In fact, regulators to-date have placed drone coordination DAA in the hand of safety pilots \cite{faaregulations} that either visually deconflict in line-of-sight operations or communicate with air traffic control by voice.

Scalable aircraft coordination requires organizational infrastructure including system-wide and vehicle-to-vehicle (V2V) datalink capable of coordinating and sharing vehicle intent (flight route). 
% Get away from government and focus on technology - you already have government mentioned in the last paragraph.  It doesn't help the paper.
%This comes in the form of governmental support, but also rules and protocols to facilitate urban navigation for drones. 
The UAS community currently envisions modest density distinct flights through sequential predefined corridors as a direct extension of today's manned airport traffic area flight routes. However, even with altitude layers, single-queue fixed corridors impose low traffic density constraints and support limited routing flexibility compared to the cooperative and deconflicted dynamic traffic routing capability proposed in this work.
%rules minimize the number of decisions to be made at the local level and allow drones to make assumptions about the shape of air traffic. The most essential of these rules is an established air network made of air corridors which lowers the complexity of urban air navigation.

We define an \texttt{air network} as an aerial analog of the multi-lane road network except that the air network is virtual so it can be dynamically reconfigured based on traffic routing requests, winds, and other factors. 
% Unlike its ground analog, an air network does not require costly physical infrastructure. 
Through datalink each UAS can update its dynamic air network map in real-time.  UAS can also store default air network structures for usage in an offline (pre-flight) mode. 

Air networks have two uses. First, they provide [closely-spaced] smooth paths that offer appropriate clearance from buildings, terrain, other structures, and neighboring UAS. The consideration environmental maps and safe separation distances significantly simplifies UAS flight planning and dynamic rerouting. Second, they define safely separated candidate paths that can assure collision avoidance so long as each UAS tracks its planned trajectory to within expected error bounds.

% \subsection{Related Work}

The authors previously presented a method to automatically generate a dense air network for any urban environment that wraps structures \cite{emadi2022finite}. This method takes as input surface elevation data including buildings and other structures and uses the fluid-inspired method presented in \cite{8907366} to generate layers of air corridors. Each layer is a fixed-altitude plane with air corridors oriented in a fixed nominal direction that wrap around buildings and other structures. 
% \subsection{Contributions and Outline}
This paper presents several advancements to the authors' previous work \cite{emadi2022finite} which increases its scalabity and usability. More specifically, this paper offers the following distinct and novel contributions:

\begin{enumerate}
    \item The acquisition method for surface elevation data has been fully automated for US cities. In our previous work \cite{emadi2022finite}, three dimensional models for buildings were manually created using Google Maps data which is not scalable. We have now built software to automatically query the United States Geological Survey (USGS)'s databases for surface elevation data in the form of point clouds \cite{nationalmap}. These point clouds are then processed to create high-resolution surface elevation maps. For cities outside the US, we use existing high-resolution 3d models of cities where this data is available to directly obtain constant-altitude "slices" of buildings and other structures.

    \item We maximize airspace usability by allowing air corridors in the urban canyons between buildings when sufficient clearance exists. This was not the case in our previous work \cite{emadi2022finite} which could only handle wrapping air corridors around a single building.

    \item We guarantee a minimum air corridor width by representing air corridors as chains of fixed-size cubes with dimensions set to assure sufficient route separation. In our previous work \cite{emadi2022finite}, air corridors were defined as the surface between two streamlines with no restrictions on inter-streamline distance. In this paper, we incrementally build corridors with chains of cubes approximating streamlines while checking at each step that the cubes do not intersect.

    \item We split the air network into connected zones where the sub-network of each zone can be generated independently. In our previous work \cite{emadi2022finite}, the air network was generated all at once by solving a differential equation. This meant that for each modification the whole network had to be regenerated from scratch. This is no longer the case. Our air network zones can be sized to manage computational cost. Solving the differential equation for a city requires a time that scales with the $3/2$ power of the city's area. When the air network is solved done for each zone independently, the computational time scales linearly with city area.
    
    \item We present sample air networks for major cities to verify that air network generation scales properly and generates dense connected networks for real cities.  Results also offer the UAS community visual examples of dense low-altitude air network routing options.

\end{enumerate}

This paper is structured as follows. Section \ref{sec:methodology} presents our method to generate an elevation map from USGS data or 3D city model data.  This section also describes our method to generate an air network from an elevation map. This method is applied to multiple major cities in Section \ref{sec:results} with a conclusion and discussion of future work in Section \ref{sec:conclusion}.

%% file: methodology.tex
\section{Methodology}
\label{sec:methodology}

This section presents the steps for generating an air network starting from raw elevation data or a 3d city model. Our method to generate air networks involves generating a discrete elevation map, solving a discrete fluid equation, and iteratively building corridors with finite-sized cubes. As such, we first describe a process to formally discretize the airspace in Section \ref{sec:discrete_space}. Then we use either USGS LiDAR elevation data or a 3d model to generate a discrete elevation map in Section \ref{sec:elevation_map}. This map is leveraged in Section \ref{sec:streamlines} to solve the ideal fluid flow equation for constant-elevation slices to obtain streamlines that wrap buildings. Finally, streamline maps are used to generate air corridors in Section \ref{sec:air_corridors}.

\input{discrete_space}

\input{elevation_map.tex}

\input{streamlines.tex}

\input{air_corridors.tex}

%% file: discrete_space.tex
\subsection{Discretization of the airspace}
\label{sec:discrete_space}

As all of the methods used in our air network generation are discrete, we first must discretize the airspace. The airspace is originally a 3-dimensional continuous space. It has a spatial origin point $\mathbf{P}_0 = (0, 0, 0)$ with a reference basis $(\vu{i}, \vu{j}, \vu{k})$.

\begin{figure}[htpb]
\centering
\includegraphics[width=3in]{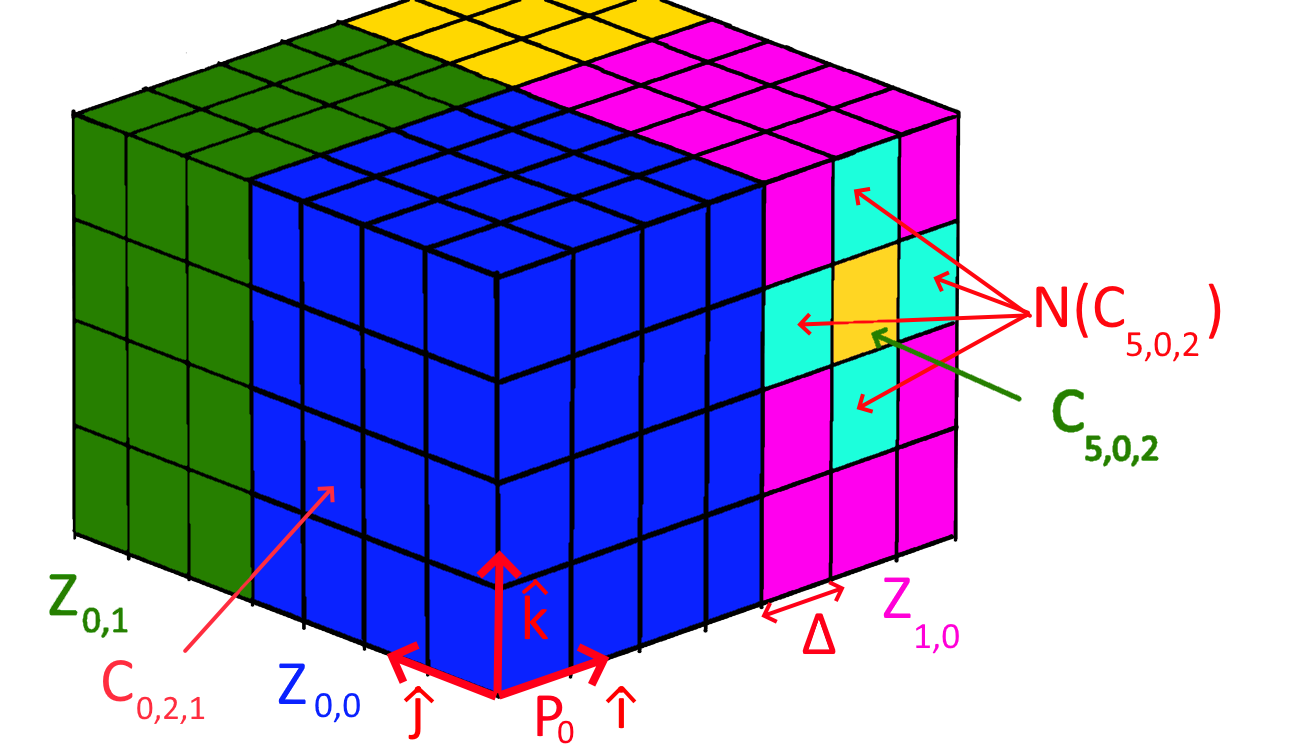}
\caption{Example cell discretization with anchor point $\mathbf{P}_0$, reference frame $(\vu{i}, \vu{j}, \vu{k})$, cell side size $\Delta$, and zone base side length $N = 4$. Zones $Z_{0,0}$, $Z_{0,1}$, and $Z_{1,0}$ are shown respectively in blue, green, and pink. Cell $C_{5,0,2}$ is shown in yellow and its neighbors $\mathbf{N}(C_{5,0,2})$ are shown in cyan.}
\label{fig:discrete_space}
\end{figure}

To discretize it, we cover the airspace with a 3-dimensional grid of cubic cells as shown in Fig. \ref{fig:discrete_space}. The axes of the grid align with the basis vectors and each cell has side length $\Delta \in \mathbb{R}^+$. We assign to each cell discrete grid coordinates $(i,j,k) \in \mathbb{Z}^3$ with each cell being referenced as $C_{i,j,k}$. The set of all cells is denoted $\mathbf{C}$. Each cell encloses a given cubic region of the originally continuous airspace. Given a point in the airspace, we can determine the cell that encloses it by getting the cell's grid coordinates using the function $\mathbf{F}$:
\begin{equation}
\begin{aligned}
\label{eq:F}
    \mathbf{F}: & \quad\ \mathbb{R}^3 & \rightarrow & \qquad\quad \  \mathbb{Z}^3 \\
    & \ (x,y,z) & \rightarrow & \left(\myfloor{\frac{x}{\Delta}}, \myfloor{\frac{y}{\Delta}}, \myfloor{\frac{z}{\Delta}}\right)
    %,\qquad \forall (x,y,z) \in \mathbb{R}^3.
\end{aligned}
\end{equation}
where $\myfloor{.}$ is the floor function.

Each cell is considered connected to the cells with which it is in direct physical contact. For this to happen, the cells' grid coordinates must differ in at least one dimension. The cells with which a cell is connected are called its Cartesian neighbors. Sample Cartesian neighbors of a cell are shown in Fig. \ref{fig:discrete_space}. The set of all Cartesian neighbors of a cell $C\in \mathbf{C}$ is denoted $\mathbf{N}(C)$ and given by function $\mathbf{N}$:
\begin{equation}
\label{eq:N}
\begin{aligned}
\mathbf{N}: & \quad\ \mathbf{C} & \rightarrow & \qquad\qquad  \mathcal{P}(\mathbf{C}) \\
    & \ \ C_{i,j,k} & \rightarrow & \{ C_{i+1,j,k}, C_{i-1,j,k},C_{i,j+1,k},   \\
 & & & \ C_{i,j-1,k}, C_{i,j,k+1}, C_{i,j,k-1}\}
\end{aligned}
\end{equation}
where power set $\mathcal{P}(\mathbf{C})$ define  the set of all subsets of $\mathbf{C}$. 

We now have a connected discretized airspace made of a grid of cells. We divide this discretized airspace into zones. This is to allow for the air network to be generated for each zone independently. Each zone consists of multiple cells. A zone has a base of $N \times N$ cells and contains all cells above them. Sample zones are shown in Fig. \ref{fig:discrete_space}. The set of all zones $\mathbf{Z}$ forms a partition of the set of all cells $\mathbf{C}$. We assign each zone coordinates $(a,b) \in \mathbb{Z}^2$ and refer to it as $\mathbf{Z}_{a,b}$. The set $\overline{\mathbf{Z}_{a,b}}$ contains the $(i,j)$ cell coordinates of the cells that $\mathbf{Z}_{a,b}$ covers. To find all the cells contained in a zone, we use the following:
\begin{subequations}
\begin{equation}
    \overline{\mathbf{Z}_{a,b}} = \left\{(i,j) \in \mathbb{Z}^2 \Bigg\vert \ 
    \begin{matrix}
    Na \leq i < N(a+1) \\ Nb \leq j < N(b+1)
    \end{matrix}\right\}, \forall (a,b) \in \mathbb{Z}^2.
\end{equation}
\begin{equation}
    \mathbf{Z}_{a,b} = \left\{C_{i,j,k}\in \mathbf{C} \big\vert \ (i,j) \in \overline{\mathbf{Z}_{a,b}}\right\}, \quad
    \forall (a,b) \in \mathbb{Z}^2.
\end{equation}
\end{subequations}

%% file: elevation_map.tex
\subsection{Elevation map generation}
\label{sec:elevation_map}

Urban airspace contains obstacles in the form of buildings and other structures. In order to generate an air network that safely wraps around all obstacles, we have to identify which cells of the discrete airspace intersect with an obstacle. As such, we split cells into two types. A free cell is a cell that does not intersect any obstacle. A full cell is a cell which does intersect an obstacle.

To simplify our model, we merge all buildings and structures with the ground and consider all of it as a terrain of varying altitude. As such, for each horizontal cell coordinates $(i,j) \in \mathbb{Z}^2$, there is a maximum cell altitude under which all cells are full and above which all cells are free. For each zone $\mathbf{Z}_{a,b} \in \mathbf{Z}$, we define a discrete elevation map $M_{a,b}: \overline{\mathbf{Z}_{a,b}} \rightarrow \mathbb{Z}$ which gives this "ground" cell altitude for each horizontal cell coordinates. As such, the type of a cell can be obtained by the function $T$:

\begin{equation}
\label{eq:type}
\begin{aligned}
T: & \quad\ \mathbf{Z}_{a,b} & \rightarrow & \qquad\quad  \{\text{full},\text{free}\} \\
    & \ \ C_{i,j,k} & \rightarrow & \begin{cases}
    \text{full}& k \leq M_{a,b}(i,j)\\
    \text{free}& k > M_{a,b}(i,j)
    \end{cases}.
\end{aligned}
\end{equation}

We use two different methods to obtain the discrete elevation map $M_{a,b}$ of a zone. If the city for which we are generating the air network is in the United States, we use the USGS's LiDAR elevation data. If the city is outside the US, we must use an existing 3d model of the city.

\subsubsection{Elevation map for US cities from USGS point clouds:}

\begin{figure}[htpb]
\centering
\includegraphics[width=3in]{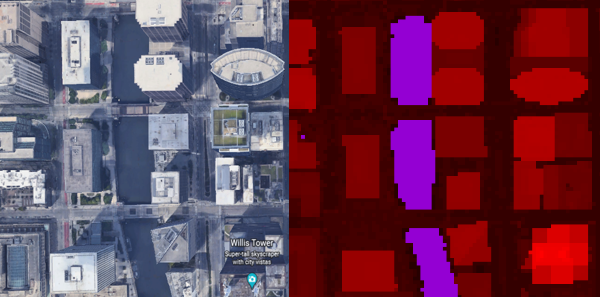}
\caption{Google Earth satellite view of city area in Chicago (left) and its corresponding elevation map  based on the USGS's LiDAR data (right). The intensity of red is proportional to the square root of cell altitude. Purple shows $(i,j)$ coordinate pairs with no altitude data where $M_{1,0}$ is positive infinity.}
\label{fig:google_earth_vs_k_map}
\end{figure}

For most regions in the US, the USGS provides LiDAR elevation data \cite{nationalmap}. 
% Not needed as this paper isn't about LiDAR --> LiDAR stands for Light Detection and Ranging. It consists in sending electromagnetic waves from a source toward an object and measuring the time it takes for the waves to reflect off the object and arrive back to the source \cite{YAN2015295}. The travel time allows to estimate the 3d position of the point of reflection. If done enough time, this process yields a set of 3d points on the surface of the object called a point cloud. 
LiDAR point clouds have been obtained from equipped aircraft and rotorcraft overflying an area to estimate terrain elevation including buildings and other structures. Note that LiDAR may not capture narrow vertical obstacles and power lines; researchers are investigating complementary methods to address this challenge \cite{flanigen2022current}. 
For a given zone, our system queries the USGS's database for any point cloud with points that are within the zone. The point clouds found are grouped into a set $\mathbf{P}_{a,b}$. The point clouds come with a projected coordinate system. So we first translate all the points such that our origin $\mathbf{P}_0$ coincides with the origin of the data's coordinate system. Then, we map the points to cells using the function $\mathbf{F}$ defined in \eqref{eq:F} to get a cell cloud $\mathbf{F}\left(\mathbf{P}_{a,b}\right)$. Finally, for each horizontal cell coordinate $(i,j)\in \overline{\mathbf{Z}_{a,b}}$, we pick the highest cell altitude from the cell cloud as the terrain altitude for the discrete elevation map $M_{a,b}(i,j)$. Therefore, $M_{a,b}(i,j)$ is given by:

\begin{equation}
\begin{aligned}
    M_{a,b}: & \quad\ \overline{\mathbf{Z}_{a,b}} & \rightarrow & \qquad\qquad  \mathbb{Z} \\
    & \quad \ (i,j) & \rightarrow & \max\left\{k\vert \ (i,j,k)\in \mathbf{F}\left(\mathbf{P}_{a,b}\right)\right\}
    %,\qquad \forall (x,y,z) \in \mathbb{R}^3.
\end{aligned}
\end{equation}

Certain horizontal coordinates might not have any cells from the cell cloud that match them. In that case, we assign infinity as the terrain altitude for them. This is the most conservative approach as it assumes the worst case scenario of there being an extremely high feature in the space with no elevation data. A sample discrete elevation map is shown in Fig. \ref{fig:google_earth_vs_k_map}.

\subsubsection{Elevation map for non-US cities from 3d models:}
In the case of non-US cities, we use exiting 3d models as our data source. For these cities, we use the 3d modeling software Blender \cite{foundation} to take slices of the 3d model at specific cell altitudes. These slices allow to distinguish between ``full'' and ``free'' cells  at every specified altitudes. 

While this method is more straightforward, it assumes the existence of a 3d model for the city which is not guaranteed. It is also for now a manual process, although it is possible to automate the slicing. Lastly, high fidelity 3d models are often paid models as opposed to USGS data which is public, free, and available for about all of the continental US.

%% file: streamlines.tex
\subsection{Streamlines map generation}
\label{sec:streamlines}

Our goal is to generate air corridors at fixed-altitude layers that wrap around buildings. For this, we solve the equation of an ideal irrotational fluid flowing in a fixed-altitude 2-dimensional plane in between buildings. This allows us to obtain the streamline distribution in the layer in question. The streamline distribution is used in Section \ref{sec:air_corridors} to guide air corridor generation.

The input parameters are the number of layers for the air network $M \in \mathbf{N}$, and for each layer $n \in \{1\cdots M\}$, its altitude $h_n$ and the nominal direction of its air corridors $\vu{d}_n$.

For each zone, we have to solve the fluid flow equation at each of the given altitudes. In this section, we fix the zone to be $\mathbf{Z}' \in \mathbf{Z}$ and the layer in question to be $M' \in \{1\cdots M\}$ such that $h' = h_{M'}$ and $\vu{d}' = \vu{d}_{M'}$.

First, we convert the continuous air space altitude $h'$ to a cell altitude $k'$ using the function $\mathbf{F}$ defined in \eqref{eq:F}. Then we take a horizontal slice $\tilde{\mathbf{Z}}$ of the zone $\mathbf{Z}'$ at cell altitude $k'$. A horizontal slice of a zone at cell altitude k is the subset of cells within the zone that have cell altitude $k$. As such:

\begin{equation}
    \tilde{\mathbf{Z}} = \left\{C_{i,j,k} \in \mathbf{Z}' \vert k = k'\right\}.
\end{equation}

If a continuous ideal irrotational fluid were to flow in the slice $\tilde{\mathbf{Z}}$, it would form streamlines such that each point in the plane would be part of a given streamline. Each streamline can be assigned a real index $\psi$, and this $\psi$ distribution satisfies Laplace's equation:
\begin{equation}
    \nabla^2\psi = 0.
\end{equation}

In our case, the slice is made of discrete cell. So the $\psi$ distribution is approximated with a discrete version which assigns to each cell in the slice $C \in \tilde{\mathbf{Z}}$ a streamline index $\psi(C)$. Laplace's equation also transform to its discrete form:
\begin{equation}
\label{eq:discrete}
\sum_{C' \in (\mathbf{N}(C)\cap \tilde{\mathbf{Z}})}\left(\psi(C) - \psi(C')\right) = 0, \quad \forall C \in \tilde{\mathbf{Z}}.
\end{equation}

In Eq. \eqref{eq:discrete}, $(\mathbf{N}(C)\cap \tilde{\mathbf{Z}})$ is the set of all neighbors of $C$ which are in the same slice. To obtain the $\psi$ distribution, we first have to define the boundary conditions for the Laplace difference equation, then solve it.

\subsubsection{Boundary conditions:}

There are two types of boundary conditions. The first is the $\psi$ values of boundary cells which control the nominal direction of streamlines. The second is the $\psi$ value of full cells which control how streamlines wrap around them.

\begin{figure}[h]
% \vspace{-.6cm}
\centering
 \subfigure[]{\includegraphics[width=0.33\linewidth]{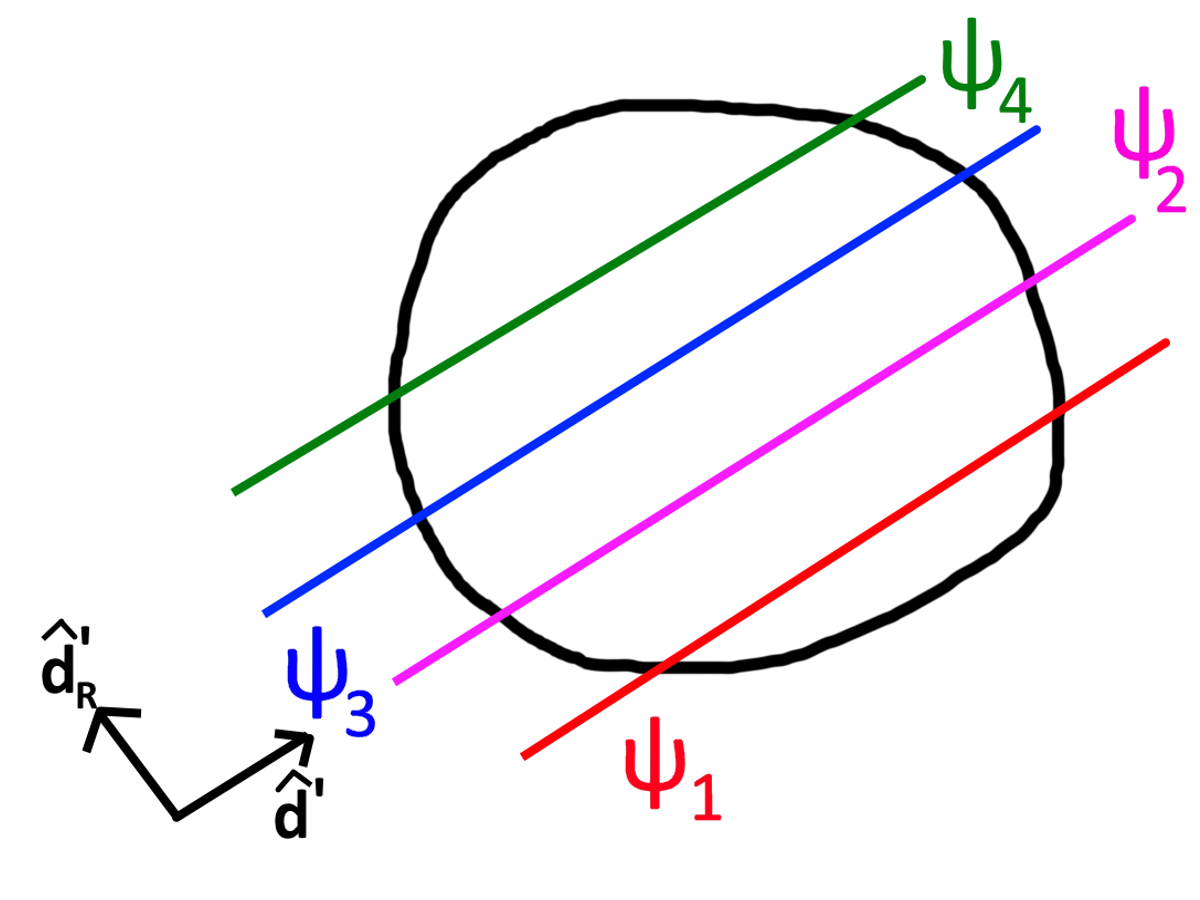}}
 \subfigure[]{\includegraphics[width=0.33\linewidth]{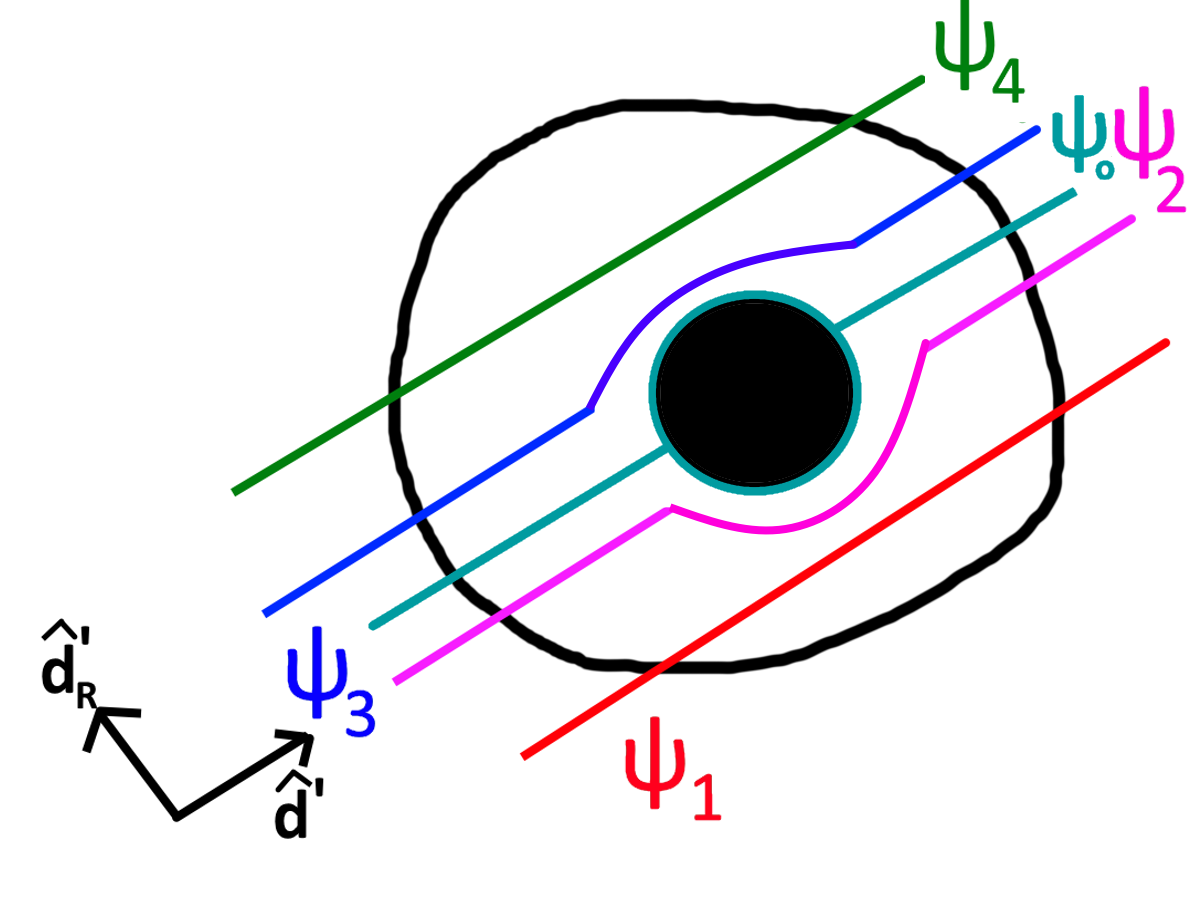}}
 \subfigure[]{\includegraphics[width=0.30\linewidth]{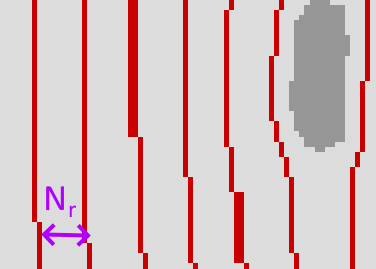}}
% \subfigure[]{\includegraphics[width=0.19\linewidth]{RegroupSnapV2.jpg}}
% \subfigure[]{\includegraphics[width=0.2\linewidth]{Hybridx.png}}
% \subfigure[]{\includegraphics[width=0.2\linewidth]{Hybrydy.png}}
% \subfigure[]{\includegraphics[width=0.81\linewidth]{FastResponse-V5.png}}
% \subfigure[]{\includegraphics[width=0.17\linewidth]{FormalSafetyFeasibility-V2.png}}
% \subfigure[]{\includegraphics[width=0.49\linewidth]{RTDK-Level.png}}
% \subfigure[]{\includegraphics[width=0.5\linewidth]{RTDCoordinateTransform.png}}
\vspace{-.3cm}
\caption{(a) Visualization of method for assigning $\psi$ values to boundary cells. The nominal streamline direction is $\vu{d}'$ while $\vu{d}'_R$ is orthogonal to it. (b) Visualization of method for assigning $\psi$ values to obstacles. The obstacle is assigned $\psi_o$ such that a single streamline passes by its boundary. Here $\psi_4 < \psi_3 < \psi_o < \psi_2 < \psi_1$. (c) Illustration of minimum cell distance between starting points of two air corridors.}
\label{fig:boundary_psia}
\end{figure}

% \begin{figure}[htpb]
% \centering
% \includegraphics[width=3in]{figures/boundary_psi.png}
% \caption{Visualization of method for assigning $\psi$ values to boundary cells. The nominal streamline direction is $\vu{d}'$ while $\vu{d}'_R$ is orthogonal to it.}
% \label{fig:boundary_psi}
% \end{figure}

The set of all cells on the boundary of the slice is denoted $\partial \tilde{\mathbf{Z}}$. Each two cells on the boundary that have the same $\psi$ value are on the same streamline. Therefore, in order for streamlines to have a nominal direction of $\vu{d}'$, any two points on a line with direction $\vu{d}'$ must have have the same $\psi$ value as shown in Fig. \ref{fig:boundary_psia} (a). One way to achieve this is to have the $\psi$ value of boundary cells change when moving orthogonally to $\vu{d}'$. The simplest way to do it is to rotate $\vu{d}'$ by 90 degrees to get ${\vu{d}'}_R$, then have the $\psi$ value of a boundary cell be the cell's coordinate along ${\vu{d}'}_R$. This yields:
\begin{equation}
\label{eq:boundary-psi}
    \psi(C_{i,j,k}) = \begin{bmatrix}i&j\end{bmatrix}^T.\vu{d}'_R, \qquad \forall C_{i,j,k} \in \partial \tilde{\mathbf{Z}},
\end{equation}
where ``$\cdot$'' is the the inner product symbol.

Eq. \eqref{eq:boundary-psi} is a more general form of the formulas used in \cite{emadi2022finite} to get the $\psi$ value of boundary cells. This is because Eq. \eqref{eq:boundary-psi} works with any nominal streamline direction $\vu{d}'$ whereas the formulas in \cite{emadi2022finite} only work for $\vu{d}' = \begin{bmatrix}1&0\end{bmatrix}^T$ or $\vu{d}' = \begin{bmatrix}0&1\end{bmatrix}^T$.

% \begin{figure}[htpb]
% \centering
% \includegraphics[width=3in]{figures/boundary_psi_2.png}
% \caption{Visualization of method for assigning $\psi$ values to obstacles. The obstacle is assigned $\psi_o$ such that a single streamline passes by its boundary. Here $\psi_4 < \psi_3 < \psi_o < \psi_2 < \psi_1$.}
% \label{fig:boundary_psi_2}
% \end{figure}

Groups of full cells that are all neighbors of each other form an obstacle. To partition full cells into distinct obstacles, we start with any non-categorized full cell and iteratively group it with full cell neighbors. Once no full cell neighbors are left, the obstacle is complete and the process starts again with another non-categorized full cell. Once all full cells have been categorized into obstacles, we assign their $\psi$ values.

To have streamlines wrap around obstacles as shown in Fig. \ref{fig:boundary_psia} (b), the boundary cells of obstacles must have the same $\psi$ value. This is equivalent to having all cells of the obstacle taking a single $\psi$ value. Streamlines with a higher $\psi$ will pass around the obstacle from one side, while streamlines with a lower $\psi$ will pass around from the other side.

Our method here differs from our previous work. In our previous paper \cite{emadi2022finite}, all obstacles were assigned $\psi=0$. This made it such that streamlines could not go in between buildings and instead all went by the boundary of the urban region. This is because setting $\psi = 0$ for all obstacles effectively groups them into a single obstacle covered by a single streamline. To fix this issue, we modified our method for assigning $\psi$ values for obstacles.

Our goal is, for each obstacle, to select the $\psi$ value which minimally disturbs the flow of streamlines. As such, we first pick the center coordinate of the obstacle $\vb{r}_o$. We then convert it to cell coordinates using the function $\mathbf{F}$. Finally, we use the same formula that we used for boundary cells, Eq. \eqref{eq:boundary-psi}, to select its $\psi$ value:
\begin{equation}
\psi_o = \mathbf{F}(\vb{r}_o)^T.\vu{d}'_R.
\end{equation}

\subsubsection{Solving the difference Laplace equation}

In this section, we introduce a new indexing system for cells in the slice $\tilde{\mathbf{Z}}$ to simplify our equations. Instead of referring to cells using three indices as in $C_{i,j,k}$, we refer to cells with a single index $I \in \{1,\cdots,m\}$ where $m=N^2$ is the number of cells in the slice. This new indexing system is designed to group cells into the following categories:
\begin{enumerate}
    \item $( 1,\cdots,m_b )$ are boundary cells.
    \item $( m_b + 1,\cdots,m_b + m_f )$ are free non-boundary cells.
    \item $( m_b + m_f + 1,\cdots,m_b + m_f + m_o )$ are full non-boundary cells.
\end{enumerate}
with:
\begin{enumerate}
    \item $\boldsymbol{\psi}_{b} = \begin{bmatrix}\psi(C_1)&\cdots&\psi(C_{m_b})\end{bmatrix}^T$.
    \item $\boldsymbol{\psi}_{f} = \begin{bmatrix}\psi(C_{m_b + 1})&\cdots&\psi(C_{m_b + m_f})\end{bmatrix}^T$.
    \item $\boldsymbol{\psi}_{o} = \begin{bmatrix} \psi(C_{m_b + m_f + 1})&\cdots&\psi(C_{m_b + m_f + m_o})\end{bmatrix}^T$.
\end{enumerate}

Let $\mathbf{L}=\left[L_{IJ}\right] \in {\mathbb{R}}^{m\times m}$ be the Laplacian matrix \cite{MERRIS1994143} for the slice $\tilde{\mathbf{Z}}$ whose entries indicate the number of neighbors each cell has in $\tilde{\mathbf{Z}}$:
\begin{equation}
    L_{IJ} = \begin{cases}
\lvert \mathbf{N}(C_I)\cap \mathbf{Z}_{k} \rvert &I = J\\
-1&I\neq J \wedge C_J \in N(C_I)\\
0&\text{otherwise}
\end{cases},
\end{equation}
where neighboring function $\mathbf{N}\left(\cdot\right)$ was defined in Eq. \eqref{eq:N}.

Matrix $\mathbf{L}$ can be split into nine sub-matrices which define the connections between the three groups of cells:
\begin{equation}\label{original}
    \mathbf{L} = \begin{bmatrix}
\mathbf{L}_{bb}&\mathbf{L}_{bf}&\mathbf{L}_{bo}\\
\mathbf{L}_{fb}&\mathbf{L}_{fc}&\mathbf{L}_{fo}\\
\mathbf{L}_{ob}&\mathbf{L}_{of}&\mathbf{L}_{oo}\\
\end{bmatrix}.
\end{equation}
With this established, we can combine all instances of the Laplace difference equation Eq. \eqref{eq:discrete} into one linear system:
\begin{equation}
\mathbf{L}_{fb}\boldsymbol{\psi}_{b} + \mathbf{L}_{fc}\boldsymbol{\psi}_{f} + \mathbf{L}_{fo}\boldsymbol{\psi}_{o} = 0.
\end{equation}

We know the $\psi$ values of all boundary and full non-boundary cells which are $\boldsymbol{\psi}_{b}$ and $\boldsymbol{\psi}_{o}$. So we can solve for the $\psi$ values of free non-boundary cells $\boldsymbol{\psi}_{f}$ with: 
\begin{equation}
\label{eq:psi-free}
\boldsymbol{\psi}_{f} = -{\mathbf{L}_{fc}}^{-1}(\mathbf{L}_{fb}\boldsymbol{\psi}_{b} + \mathbf{L}_{fo}\boldsymbol{\psi}_{o}).
\end{equation}

 This linear system should not be solved with a regular solver. This is because $\mathbf{L}$ is usually very large as its size is $m^2 = N^4$. Also the number of its non-zero entries is in the order of $N^2$, 
which makes it a sparse matrix \cite{davis_rajamanickam_sid-lakhdar_2016}. As such, we using a sparse linear system solving method to solve the set of linear equations provided by Eq. \eqref{original} with a low computation cost. Because both $\mathbf{L}$ and $\mathbf{L}_{fc}$ are symmetric, we use the conjugate gradient method \cite{cg} to obtain $\boldsymbol{\psi}_{f}$.
% As beings neighbors is a symmetric property, $\mathbf{L}$ is symmetric and so $\mathbf{L}_{fc}$ is also symmetric. This allows using the conjugate gradient method \cite{cg} to solve for $\boldsymbol{\psi}_{f}$.

%% file: air_corridors.tex
\subsection{Air corridor generation}
\label{sec:air_corridors}

With a complete discrete streamline distribution $\psi$, we can finally generate air corridors. Air corridors are chains of connected cells that start and end on the boundary of the slice $\Tilde{\mathbf{Z}}$. They also approximate streamlines and do not intersect obstacles or each other.

Our definition of air corridors ensures they have a finite width equal to the size of cells $\Delta$. This differs from previous work \cite{emadi2022finite} where air corridors were defined as the area between two streamlines. The problem with directly using streamlines to get corridor boundaries is that we could not find a straightforward method to ensure streamlines are separated by a minimum distance. This is important as drones have a finite width, so we must ensure that air corridors have a minimum width which is what our new definition of air corridors does.

To generate an air corridor, we start at a boundary cell $C_b \in \partial\Tilde{\mathbf{Z}}$. Boundary cells are split into three groups based on a property we call orientation.
\begin{enumerate}
    \item If at $C_b$'s position, $\vu{d}'$ points inside the slice, $C_b$ is oriented forward.
    \item If at $C_b$'s position, $\vu{d}'$ points outside the slice, $C_b$ is oriented backwards.
    \item Otherwise, the $C_b$ has no orientation.
\end{enumerate}

We only consider $C_b$ as a starting point for a corridor if it has an orientation. If $C_b$ is orientated forward, we expand the corridor in the $\vu{d} = \vu{d}'$ direction. If $C_b$ is oriented backwards, we use $\vu{d} = -\vu{d}'$ instead. To generate the air corridor, we use algorithm \ref{alg:corridor}.

\RestyleAlgo{ruled}

\begin{algorithm}
\caption{Algorithm to generate a corridor from the $\psi$ distribution.}
\KwData{$C_b$, $\vu{d}$, $\psi$, previously built air corridors}
\KwResult{AirCorridor is an air corridor that starts at $C_b$ and goes in the $\vu{d}$ direction}
 LastCell $\gets C_b$\;
 AirCorridor $\gets$ List\;
 Add LastCell to AirCorridor\;
OriginalPsi $\gets \psi(C_b)$\; 
\While{LastCell $\notin \partial \mathbf{Z}_{k}$ OR Orientation(LastCell) = Orientation($C_b$)}{
    Neighbors $\gets N($LastCell$) \cap \mathbf{Z}_{k}$\;
    Discard neighbors that are full\;
    Discard neighbors that are part of AirCorridor\;
    Discard neighbors whose progress along $\vu{d}$ is smaller than the progress of LastCell\;
    (The progress of a cell $C_{i,j,k} \in \mathbf{Z}_{k}$ along $\vu{d}$ is defined as $\begin{bmatrix}i & j\end{bmatrix}^T \cdot \vu{d}$.)\;
    
    \If{No neighbors remain}{
        Stop, attempt failed\;
    }
    
    NextCell $\gets$ neighbor whose $\psi$ is the closest to OriginalPsi\;
    
    \If{NextCell \textit{is part of another corridor}}{
        Stop, attempt failed\;
    }
  
    Add NextCell to AirCorridor\;
    LastCell $\gets$ NextCell\;
}
Attempt successful\;

\label{alg:corridor}
\end{algorithm}

The main idea behind algorithm \ref{alg:corridor} is to iteratively add cells while trying to follow the streamline. the air corridor is a list of cells. At each run through the loop, all neighbors of the last cell in the corridor are selected. From the selected cells, we remove the cells that are full, already part of the corridor, or that would not advance the corridor along the direction of expansion $\vu{d}$. This is to ensure the corridor does not contain obstacle cells and does not loop too much. If none of them remain, then a dead end was reached and the corridor is discarded. Otherwise, we pick the neighbor with the closest $\psi$ value to that of the original cell $C_b$. This is to ensure the corridor stays near the same streamline. If the selected cell is already part of established corridor, it is considered a dead end and the corridor is discarded. This is to avoid corridor intersections and excessive deviations. If the selected cell is available, it is added to the corridor and the loop runs again. The loop stops when the corridor reaches a boundary cell with a different orientation from $C_b$, meaning it reached the other end of the slice boundary.

Air corridor generation is attempted at each cell on the boundary of the slice. The zone for which we are generating these air corridors might have neighbor zones with existing sub-networks. If that is the case, we give priority to the boundary cells which would allow connections with the neighbor zones' existing corridors. When all of these potential connections are attempted, then we try the rest of the boundary cells. This ensures maximum inter-zone connectivity for the air network.

% \begin{figure}[htpb]
% \centering
% \includegraphics[width=3in]{figures/Nr.png}
% \caption{Illustration of minimum cell distance between starting points of two air corridors.}
% \label{fig:Nr}
% \end{figure}

To avoid high variations in the network density, we add a minimum cell distance $N_r$ between boundary cells at which we attempt to generate corridors (See Fig. \ref{fig:boundary_psia}(c)).

Within the zone $\mathbf{Z}'$, for each two vertically adjacent layers of corridors, we allow vertical connections between them. These vertical connections are in the form of vertical chains of cells which connect the corridors of each of the two layers. They happen when cells with the same horizontal coordinate $(i,j)$ within each slice are both part of a corridor in their respective layer. This ensures maximum inter-layer connectivity for the air network.

%% file: results.tex
\section{Results}
\label{sec:results}

\begin{figure*}[h]
% \vspace{-.6cm}
\centering
 \subfigure[]{\includegraphics[width=0.26\linewidth]{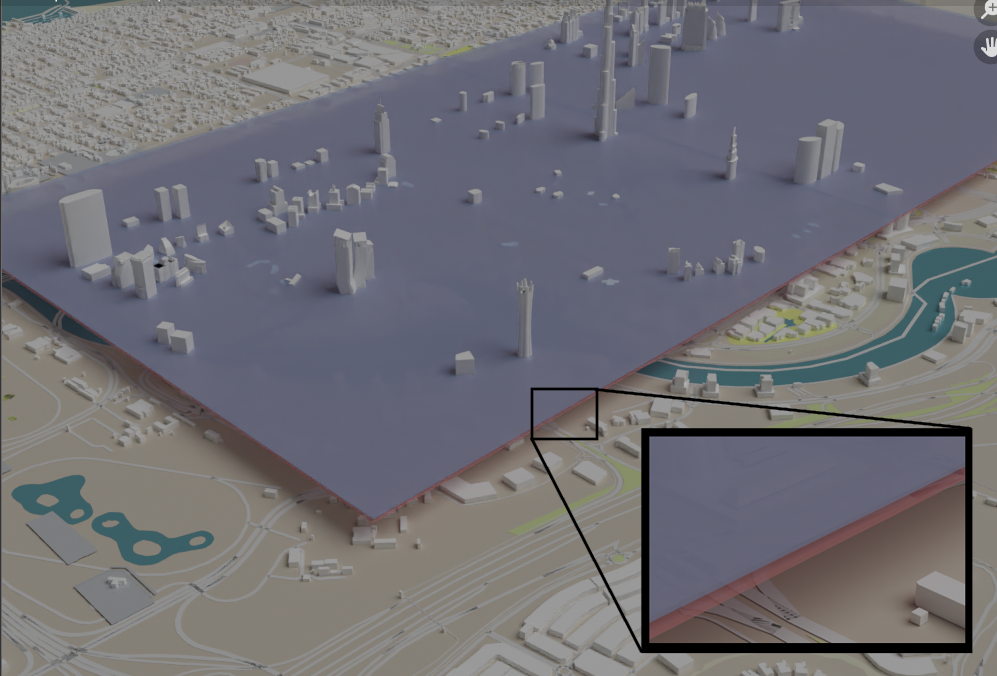}}
 \subfigure[]{\includegraphics[width=0.27\linewidth]{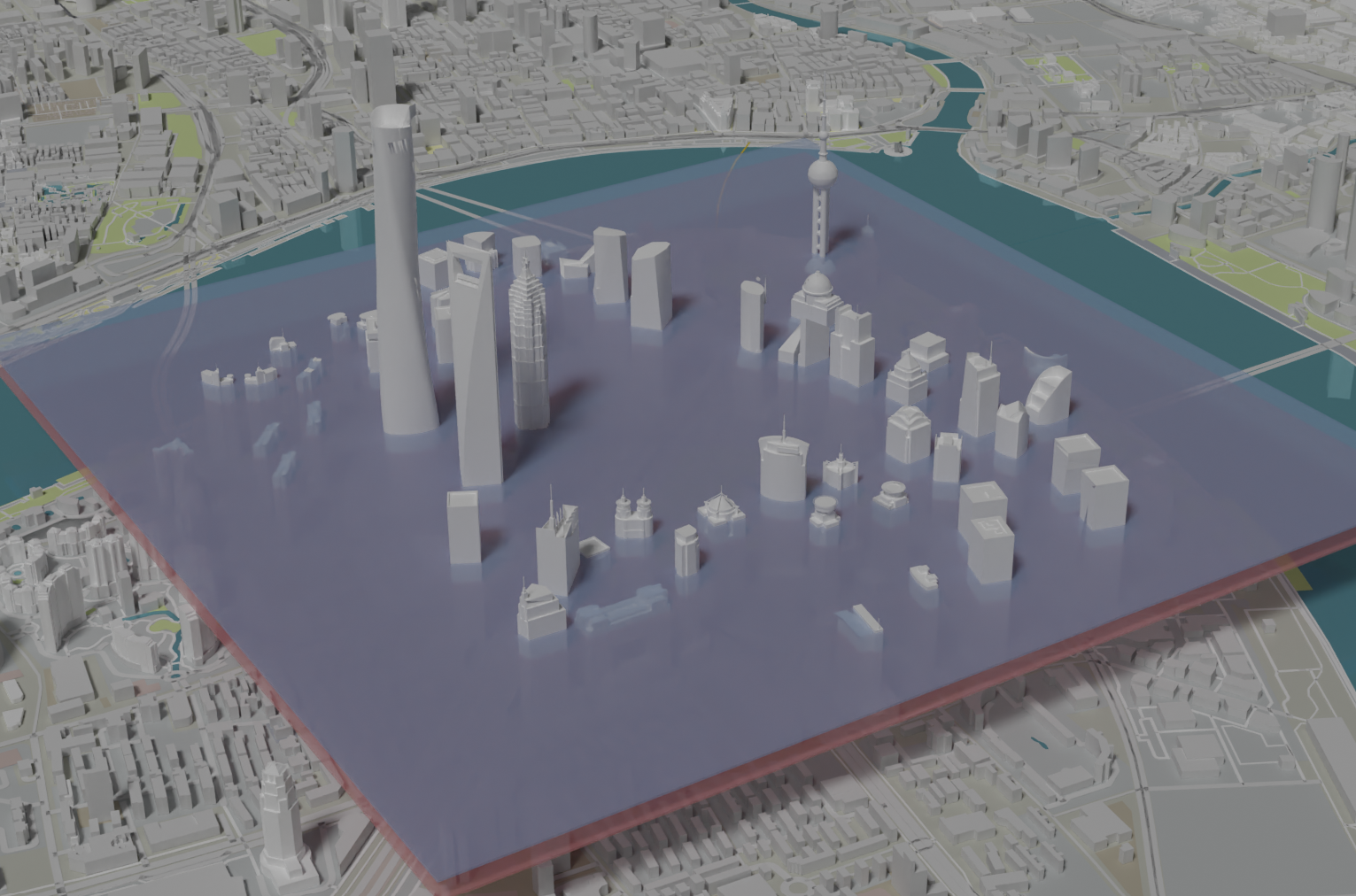}}
  \subfigure[]{\includegraphics[width=0.27\linewidth]{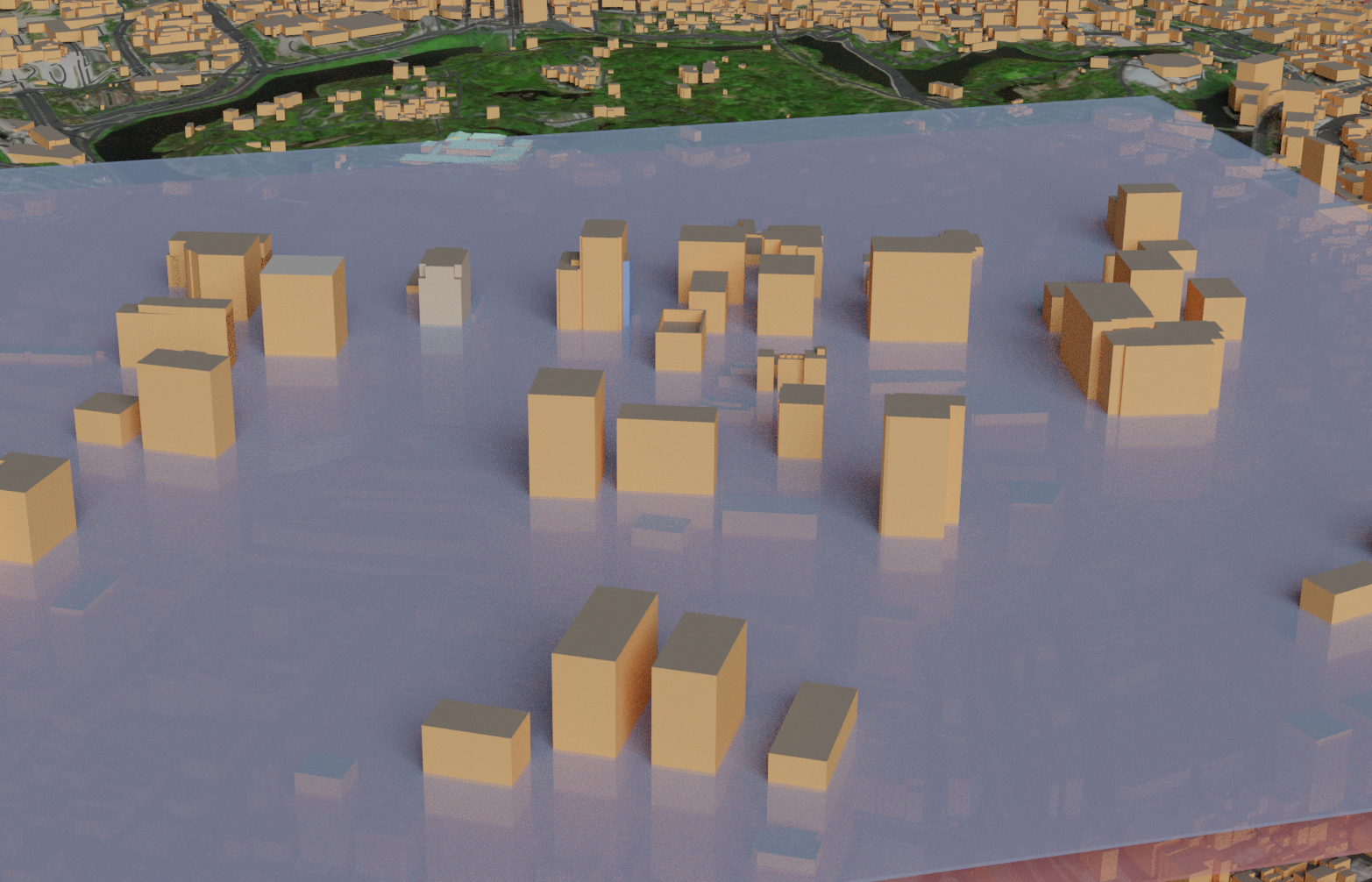}}
 \subfigure[]{\includegraphics[width=0.18\linewidth]{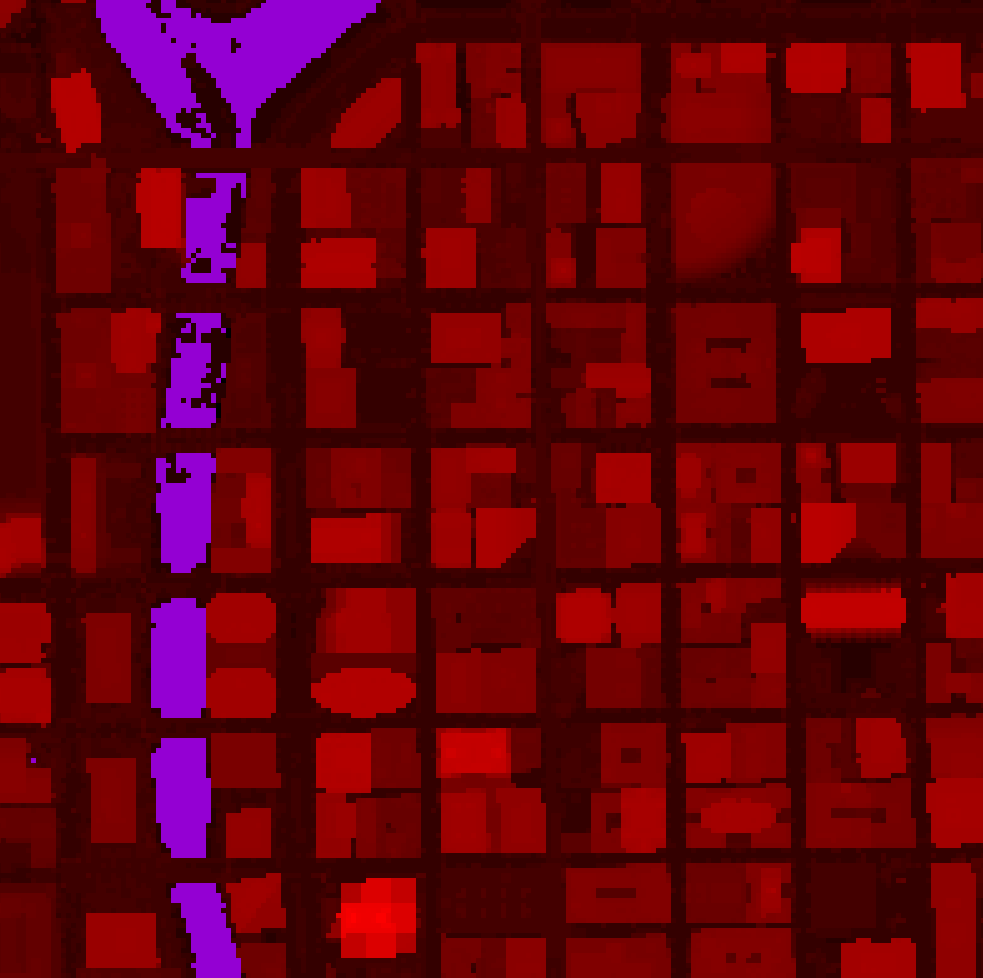}}
% \subfigure[]{\includegraphics[width=0.20\linewidth]{batteryschematic.png}}
% \subfigure[]{\includegraphics[width=0.19\linewidth]{RegroupSnapV2.jpg}}
% \subfigure[]{\includegraphics[width=0.2\linewidth]{Hybridx.png}}
% \subfigure[]{\includegraphics[width=0.2\linewidth]{Hybrydy.png}}
% \subfigure[]{\includegraphics[width=0.81\linewidth]{FastResponse-V5.png}}
% \subfigure[]{\includegraphics[width=0.17\linewidth]{FormalSafetyFeasibility-V2.png}}
% \subfigure[]{\includegraphics[width=0.49\linewidth]{RTDK-Level.png}}
% \subfigure[]{\includegraphics[width=0.5\linewidth]{RTDCoordinateTransform.png}}
\vspace{-.3cm}
\caption{3d models used for (a) Dubai \cite{cgtrader_dubai}, Shanghai \cite{cgtrader_shanghai}, and  Tokyo \cite{cgtrader_tokyo}. In these  three sub-figures, the slice for the lower layer of the air network is shown in red while the slice for the upper layer of the air network is shown in blue. (d) Discrete elevation map $M_{1,0}$ of the zone $\mathbf{Z}_{1,0}$ for Chicago based on the USGS's LiDAR data. The intensity of red is proportional to the square root of cell altitude. Purple shows $(i,j)$ coordinate pairs with no altitude data where $M_{1,0}$ is positive infinity.}
\label{maps}
\end{figure*}

We tested our air network generation method on sections of the cities listed in Table \ref{tab:cities}. Table \ref{tab:cities} also includes the corresponding air network parameters table for each city and the time it took to generate them. For Dubai, Shanghai, and Tokyo, we used existing 3d models. For Chicago, we used USGS LiDAR elevation data to create their discrete elevation maps. 
Pictures of the 3d models for ubai, Shanghai, and Tokyo are shown in Fig. \ref{maps}(a), Fig. \ref{maps}(b), and Fig. \ref{maps}(c) respectively. These figures also show the constant-altitude slices made for the air network layers. The discrete elevation maps $M_{1,0}$ of the zone $\mathbf{Z}_{1,0}$ for Chicago is shown in Fig. \ref{maps}(d).

\begin{figure*}[htpb]
\centering
\includegraphics[width=6.5in]{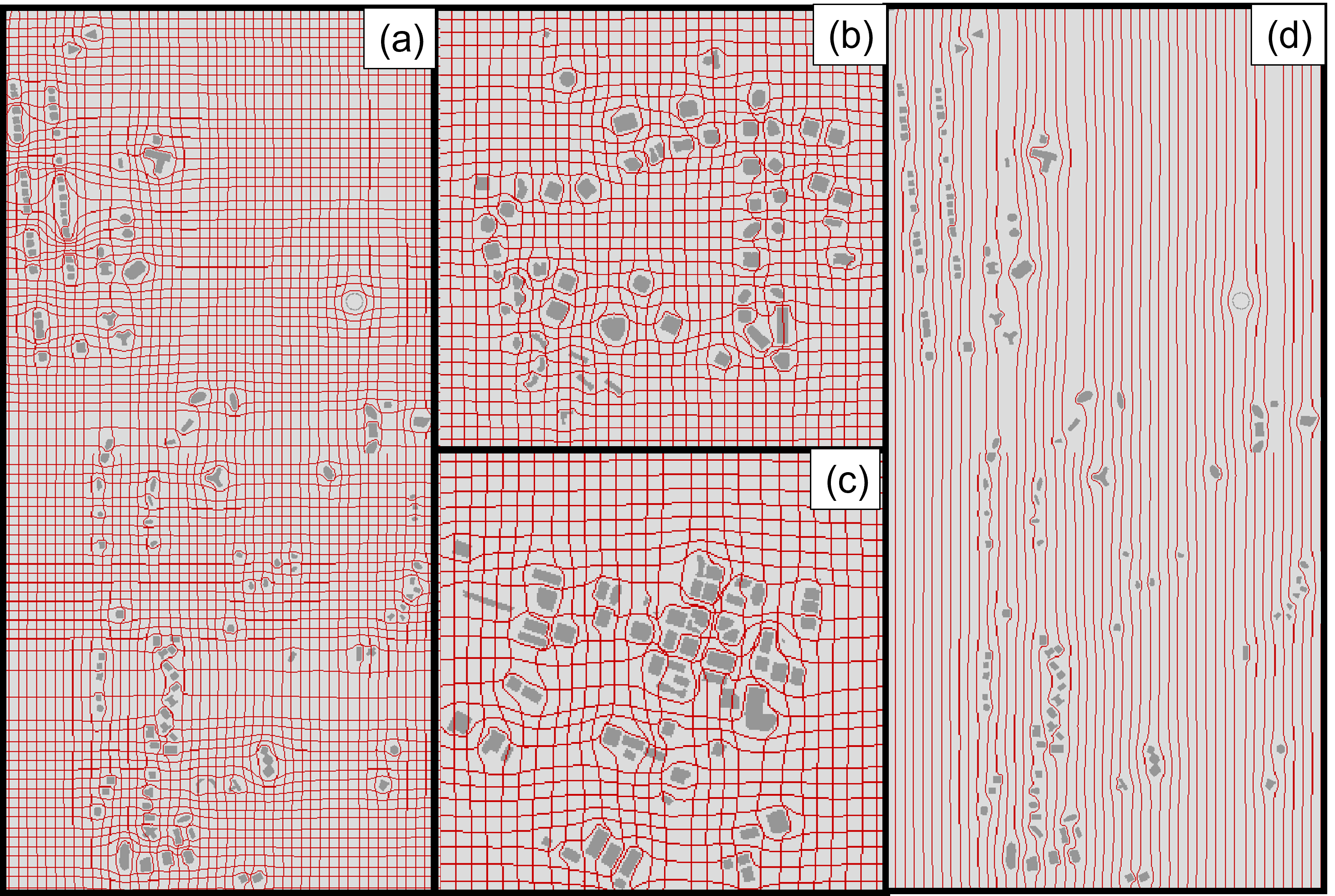}
\caption{Air network for (a) Dubai, (b) Shanghai, and (Tokyo) through zones $\mathbf{Z}_{0,0}$ (bottom) and $\mathbf{Z}_{0,1}$ (top). In these sub-figures (a-c), red shows corridor cells in both layers while grey shows full cells through the slice at the lower layer's altitude. (d) Upper layer of air network for Dubai through zones $\mathbf{Z}_{0,0}$ (bottom) and $\mathbf{Z}_{0,1}$ (top). Red shows corridor cells while grey shows full cells through the slice at the upper layer's altitude.}
\label{3dmodelcities}
\end{figure*}

\begin{figure}[htpb]
\centering
\includegraphics[width=3in]{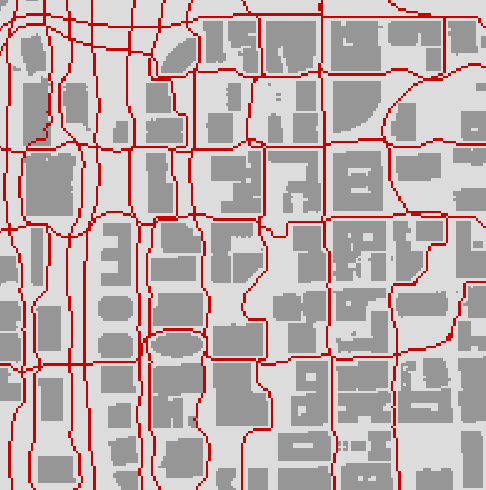}
\caption{Air network for Chicago through zones $\mathbf{Z}_{0,0}$ (bottom) and $\mathbf{Z}_{0,1}$ (top). Red shows corridor cells in both layers while grey shows full cells through the slice at the lower layer's altitude.}
\label{fig:chicago_network}
\end{figure}

For all cities, the air network was made to have two layers with orthogonal nominal flow directions. The lower level flows in the +x direction while the upper level flows in the +y direction. The generated air networks for Dubai, Shanghai, and Tokyo are shown in Fig. \ref{3dmodelcities} (a), Fig. \ref{3dmodelcities} (b), and Fig. \ref{3dmodelcities} (c) respectively. As seen in Figs. \ref{3dmodelcities} (a-c), the generated air network is composed of two zones $\mathbf{Z}_{0,0}$ and $\mathbf{Z}_{0,1}$.  Figure \ref{3dmodelcities} (d)  shows the upper layer of Dubai's air network which show cases that air corridors do not intersect (The upper layer of Dubai's air network demonstrates that our method successfully generates inter-zone connections for a high percentage of corridors). Furthermore, the air network generated for Chicago is shown in Fig. \ref{fig:chicago_network}. We notice that the air network figures (Figs. \ref{3dmodelcities} (a-c) and Fig. \ref{fig:chicago_network}) show the networks as seen from above, so the air corridors appear to intersect. This is not the case since corridors flowing in +x and corridors flowing in +y are in two different layers separated vertically.   

\begin{table}[htbp]
\centering
\caption{Cities used to get sample air networks}
  \begin{threeparttable}[t]
  \centering
       \begin{tabular}{lll}
    \toprule
     City, Country & Parameters & Generation time\\
    \midrule
    Dubai, United Arab Emirates & Table \ref{tab:dubai_param} & 20.0 s\\
    Shanghai, China & Table \ref{tab:shanghai_param} & 3.63 s\\
    Tokyo, Japan & Table \ref{tab:tokyo_param} & 3.00 s\\
    Chicago, United States & Table \ref{tab:chicago_param} & 1.78 s\\
     \bottomrule
  \end{tabular}
    \end{threeparttable}
    \label{tab:cities}
\end{table}

\begin{table}[htbp]
\centering
\caption{Parameters of air network for Dubai}
  \begin{threeparttable}[t]
  \centering
       \begin{tabular}{lll}
    \toprule
     Name&Value\\
    \midrule
        Anchor $\mathbf{P}_0$ longitude & lon = 55.251389°\\
        Anchor $\mathbf{P}_0$ latitude & lat = 25.187778°\\
        Anchor $\mathbf{P}_0$ AMSL altitude & alt = 2 meters\\
        Coordinate system & Custom\\
        Cell size & $\Delta$ = 5 meters\\
        Zone horizontal size & N = 476\\
        First corridor layer AMSL altitude & $h_1$ = 80 meters\\
        First corridor layer direction & $\vu{d}_1 = \begin{bmatrix}1.0& 0.0\end{bmatrix}$\\
        Second corridor layer AMSL altitude & $h_2$ = 90 meters\\
        Second corridor layer direction & $\vu{d}_2 = \begin{bmatrix}0.0& 1.0\end{bmatrix}$\\
        Minimum cells between ends of corridors & $N_r$ = 10\\
     \bottomrule
  \end{tabular}
    \end{threeparttable}
    \label{tab:dubai_param}
\end{table}

\begin{table}[htbp]
\centering
\caption{Parameters of air network for Shanghai}
  \begin{threeparttable}[t]
  \centering
       \begin{tabular}{lll}
    \toprule
     Name&Value\\
    \midrule
        Anchor $\mathbf{P}_0$ longitude & lon = 121.496944°\\
        Anchor $\mathbf{P}_0$ latitude & lat = 31.22889°\\
        Anchor $\mathbf{P}_0$ AMSL altitude & alt = 0 meters\\
        Coordinate system & Custom\\
        Cell size & $\Delta$ = 5 meters\\
        Zone horizontal size & N = 350\\
        First corridor layer AMSL altitude & $h_1$ = 99 meters\\
        First corridor layer direction & $\vu{d}_1 = \begin{bmatrix}1.0& 0.0\end{bmatrix}$\\
        Second corridor layer AMSL altitude & $h_2$ = 129 meters\\
        Second corridor layer direction & $\vu{d}_2 = \begin{bmatrix}0.0& 1.0\end{bmatrix}$\\
        Minimum cells between ends of corridors & $N_r$ = 10\\
     \bottomrule
  \end{tabular}
    \end{threeparttable}
    \label{tab:shanghai_param}
\end{table}

\begin{table}[htbp]
\centering
\caption{Parameters of air network for Tokyo}
  \begin{threeparttable}[t]
  \centering
       \begin{tabular}{lll}
    \toprule
     Name&Value\\
    \midrule
        Anchor $\mathbf{P}_0$ longitude & lon = 139.776944°\\
        Anchor $\mathbf{P}_0$ latitude & lat = 35.675278°\\
        Anchor $\mathbf{P}_0$ AMSL altitude & alt = 4 meters\\
        Coordinate system & Custom\\
        Cell size & $\Delta$ = 5 meters\\
        Zone horizontal size & N = 287\\
        First corridor layer AMSL altitude & $h_1$ = 76 meters\\
        First corridor layer direction & $\vu{d}_1 = \begin{bmatrix}1.0& 0.0\end{bmatrix}$\\
        Second corridor layer AMSL altitude & $h_2$ = 106 meters\\
        Second corridor layer direction & $\vu{d}_2 = \begin{bmatrix}0.0& 1.0\end{bmatrix}$\\
        Minimum cells between ends of corridors & $N_r$ = 10\\
     \bottomrule
  \end{tabular}
    \end{threeparttable}
    \label{tab:tokyo_param}
\end{table}

\begin{table}[htbp]
\centering
\caption{Parameters of air network for Chicago}
  \begin{threeparttable}[t]
  \centering
       \begin{tabular}{lll}
    \toprule
     Name&Value\\
    \midrule
        Anchor $\mathbf{P}_0$ longitude & lon = -87.65250°\\
        Anchor $\mathbf{P}_0$ latitude & lat = 41.87833°\\
        Anchor $\mathbf{P}_0$ AMSL altitude & alt = 4 meters\\
        Coordinate system & EPSG 6455\\
        Cell size & $\Delta$ = 5 meters\\
        Zone horizontal size & N = 200\\
        First corridor layer AMSL altitude & $h_1$ = 235 meters\\
        First corridor layer direction & $\vu{d}_1 = \begin{bmatrix}1.0& 0.0\end{bmatrix}$\\
        Second corridor layer AMSL altitude & $h_2$ = 245 meters\\
        Second corridor layer direction & $\vu{d}_2 = \begin{bmatrix}0.0& 1.0\end{bmatrix}$\\
        Minimum cells between ends of corridors & $N_r$ = 10\\
     \bottomrule
  \end{tabular}
    \end{threeparttable}
    \label{tab:chicago_param}
\end{table}

% \begin{figure}[htpb]
% \centering
% \includegraphics[width=3in]{figures/dubai_network_single.png}
% \caption{Upper layer of air network for Dubai through zones $\mathbf{Z}_{0,0}$ (bottom) and $\mathbf{Z}_{0,1}$ (top). Red shows corridor cells while grey shows full cells through the slice at the upper layer's altitude.}
% \label{fig:dubai_network_single}
% \end{figure}

%% file: conclusion.tex
\section{Conclusion}
\label{sec:conclusion}

This paper presented an an air network generation method with potential to safely support high UAS traffic densities for low-altitude flights through dense urban regions. 
 This work improves on the authors' previous work in terms of automation, scalability, safety of route separation, and computational overhead.  Fluid flow equations used to define streamlines between buildings was solved using a sparse linear system solver and solved independently for each zone of the city to minimize computation time. Air corridors were guaranteed to have a minimum width through the use of iterative generation with finite-sized blocks.

Future research should focus on improving the process more by accounting for variance in altitude.  Impact of air network parameters such as zone size, number of layers, and density should also be further examined in the context of metrics including but not limited to safe separation, traffic density, datalink delay, and realistic UAS trajectory tracking accuracies. 
 Future work might also examine how this work can evolve to safely and efficiently support UAS with different performance characteristics operating in the same urban air network.